# Security Awareness and Affective Feedback: Categorical Behaviour vs. Reported Behaviour

Lynsay A. Shepherd, Jacqueline Archibald
School of Arts, Media and Computer Games
Abertay University
Dundee, Scotland
lynsay.shepherd@abertay.ac.uk, j.archibald@abertay.ac.uk

*Abstract*— **A lack of awareness surrounding secure online behaviour can lead to end-users, and their personal details becoming vulnerable to compromise. This paper describes an ongoing research project in the field of usable security, examining the relationship between end-user-security behaviour, and the use of affective feedback to educate end-users. Part of the aforementioned research project considers the link between categorical information users reveal about themselves online, and the information users believe, or report that they have revealed online. The experimental results confirm a disparity between information revealed, and what users think they have revealed, highlighting a deficit in security awareness. Results gained in relation to the affective feedback delivered are mixed, indicating limited short-term impact. Future work seeks to perform a long-term study, with the view that positive behavioural changes may be reflected in the results as end-users become more knowledgeable about security awareness.**

## I. Introduction

Risky security behaviour displayed by end-users has the potential to leave devices vulnerable to compromise [1]. Despite the availability of security tools such as firewalls and virus scanners, designed to aid users in defending themselves against online threats, these tools cannot stop users engaging in risky behaviour in the context of a browser-based environment. This indicates a need to assess the current behaviour of end-users, and to educate them regarding the security implications of their actions online. Previous research into educational tools suggest the use of affective feedback as a possible method to utilise in a browser-based environment.

As part of a research project, a prototype Firefox extension named Spengler-Zuul was developed, monitoring user actions, and employing the use of affective feedback as a potential method of user education. This paper outlines part of a research project whereby a series of experiments have been conducted to gauge how behaviour logged by the aforementioned tool (categorical information) compares to behaviour reported in follow-up questionnaires with the users. Providing a comparison highlights levels of security awareness in end-users, and aids in demonstrating the potential role affective feedback can have in security education.

## II. Background

This section will outline risky security behaviours users may encounter when browsing the web. It discusses studies covering methods of measuring risk perception. Owing to the reliance on the internet, several pieces of research have posited the need to educate end-users regarding security behaviour, highlighting areas in which they may be vulnerable online. A number of existing tools are reviewed, prior to a discussion of the role of affective feedback in an educational environment. A novel approach utilising a combined affective feedback and monitoring solution is described, before the disparity between categorical user behaviour versus reported user behaviour is explored.

### A. Risky security behaviour

What constitutes risky behaviour is not necessarily obvious to all end-users and therefore, it can be difficult to recognise. Examples of such behaviour can include: interacting with a website containing coding vulnerabilities [2], downloading data from unsafe websites [3] or, creating weak passwords/sharing passwords with colleagues [4][5].

A number of studies have been conducted, in an attempt to define and categorise risky security behaviour. In 2012, a taxonomy was developed by Padayachee [6] to categorise compliant security behaviours whilst investigating if particular users had a predisposition to adhering to security behaviour. The results of the research highlighted elements which may influence security behaviours in users e.g. extrinsic motivation, identification, awareness and organisational commitment.

Another study was documented in a 2005 paper by Stanton et al [4]. Interviews were conducted with IT and security experts, in addition to a study involving end-users in the US, across a range of professions. The findings produced a taxonomy consisting of 6 identified risky behaviours: intentional destruction, detrimental misuse, dangerous tinkering, naïve mistakes, aware assurance and basic hygiene.

Milne et. al. [9] investigated risky behaviours in relation to self-efficacy. During the survey, participants were asked if they had engaged in specific risky behaviours online. These suggestions were drawn from previous research into risky behaviours [10-11]. The paper concludes that depending on the demographic and the self-efficacy of the end-user, different types of behaviour are exhibited online.

Behaviours users were asked about in the survey included the use of private email addresses to register for contests on websites, selecting passwords consisting of dictionary words, and accepting unknown friends on social networking sites. The most common risky behaviour which participants admitted to was allowing the computer to save passwords: 56% of participants admitted to this.

A lack of perception regarding online security risks can leave users, and their devices vulnerable to compromise.

*B. Measuring perception of risk*

Over the years, a variety of techniques have been utilised in an attempt to measure the perception of risk which the end-user possesses. Hill and Donaldson proposed a methodology to integrate models of behaviour and perception [10].The research attempted to assess the perception of security the system administrator possesses, and create a trust model which reduces the threat from malicious software. The methodology engaged system administrators whilst developing the threat modelling process, and quantified risk of threats, essentially creating a triage system to deal with issues.

In a different scenario, Ur et. al. [11] investigated the correlation between users' perceptions of password strengths and their actual strength on smartphones. The research employed the use of an online study to measure users thought on password strength and memorability, and their understanding of potential attacks. This data was compared against to users' perceptions regarding how passwords would fare against password cracking attacks. Comparing the data, allowed for the perception of risky behaviours to be determined.

Ng, Kankanhalli and Xu [12] devised a health belief analogy when explaining the perception of risk in terms of cyber security. Experiments were conducted with an example based upon email attachments. It was concluded that users' security behaviour could be determined via perceived susceptibility, perceived benefits, and self-efficacy.

San-José and Rodriguez [13] used a multimodal approach to measure perception of risk. In a study of over 3000 households with PCs connected to the internet, users were given an antivirus program to install which scanned the machines on a monthly basis. The software was supplemented by quarterly questionnaires, allowing levels of perception to be measured and compared with virus scan results. Users were successfully monitored and results showed that the antivirus software created a false sense of security and that users were unaware of how serious certain risks could be.

*C. Education and awareness of risky security behaviours*

A variety of tools have been developed to address differing aspects of risky security behaviours, and these are outlined in this section.

One such example is the password strength meters used in research by Ur et. al [14]. These meters were placed next to password fields and improved the security and usability of passwords. The tool was deemed to be a useful aid in password creation with participants noting that use of words such as "weak" encouraged them into creating a stronger password. However, there were potential issues with retention, and 38% of participants admitted to writing down their password from the previous day.

Other research has explored educating users regarding phishing attempts. Such tools have included Anti-Phishing Phil by Sheng et. al [15] which attempt to gamify the subject. After playing the game, 41% of participants viewed the URL of the web page, checking if it was genuine. The game produced some unwanted results in that participants became overly cautious, producing a number of false-positives during the experimental phase. PhishGuru is another training tool designed by Kumaraguru et. al [16] to discourage people from revealing information in phishing attacks. When a user clicks on a link in a suspicious email, they are presented with a cartoon message, warning them of the dangers of phishing. In a short-term study, the cartoon proved to be effective: participants retained the information after 28 days.

A newer tool, NoPhish has been developed to educate users about phishing on mobile devices [17]. The game features multiple levels where users are presented with a URL and are asked if is a legitimate link or a phishing attempt. Participants gave significantly more correct answers when asked about phishing after playing the game. In a longer-term study results showed participants still performed well however, their overall performance decreased.

Information that allows phishing emails to be targeted towards specific users can come from revealing too much information online. A proposed series of nutrition labels for online privacy have been designed in an effort to reduce risky behaviour [18]. The nutrition labels seek to present the information in an easily readable format, aiding users to understand privacy policies online. Results from a small study found that visually, the labels were more interesting to read than a traditional security policy and presented an easier way for users to find information.

A Firefox extension developed by Maurer [19] attempts to provide alert dialogs when users are entering sensitive data such as credit card information. The extension seeks to raise security awareness, providing large JavaScript dialogs to warn users, noting that the use of certain colours made the user feel more secure.

More recently, Volkamer et. al. developed a Firefox Add-On, called PassSec in attempt to help users detect websites which provided insecure environments for entering a password [20]. The extension successfully raised security awareness and significantly reduced the number of insecure logins.

The tools discussed in this section span a number of years, and some of the research may seem outdated. However, the range, and age of the research tools developed indicates there is still a problem with effectively educating users regarding security awareness. This suggests a different approach is required for user education: the use of affective feedback is a potential approach.

*D. Affective feedback and risky behaviours*

Affective feedback is defined as *"the process of using technology to help people achieve and maintain specific internal states"* [21] i.e. using signals to alter user behaviour. Previous research has indicated affective feedback may serve as a successful method of educating users about risky security behaviours [21][22][23]. User's attitudes regarding risky security behaviour must be modified in a bid to keep them safer online. Thus, by influencing end-users via affective feedback it may be possible to positively impact upon the security awareness of the end-user.

Virtual human characters, avatars, and textual content [24] and the use of colour and sound [21] have been used to influence state. Avatars provide affective feedback and have been seen to be beneficial in educational environments [21][22][23]. Textual information with the use of specific words also has the potential to alter a user's state/behaviour e.g. a password may be described as "weak" and this can encourage them to create a stronger password [14]. Colour is also often utilised, with green or blue used to imply a positive occurrence, with red indicating a negative outcome [14].

To further the argument for use of affective feedback Wixon [25] discusses its benefits but also calls for more studies into the role of affective computing, placing emphasis on the need for empirical data. This is an argument also put forward by Beale and Creed [26] in their overview of emotional simulations. Affective feedback has the potential to be utilised in the field of security education, thus the application of such a mechanism in this research project.

Research conducted in the following section seeks to utilise an affective approach, deploying the use of a monitoring solution with an integrated affective feedback delivery system, in an attempt to improve end-user security awareness.

## III. METHODOLOGY

*A. Prototypes developed*

A XUL-based Firefox extension was developed for the research project, named Spengler-Zuul. This incorporated a monitoring solution capable of detecting potential security risks such as if a page contains malicious links, or a password entered it too short. These risky behaviours were drawn from previous research [7][8][9] and were chosen as they could apply to the context of a browser-based environment. When the user interacts with the browser, the information is encrypted, and processed on the server. As an example, processing the information on a server allows the URL of a current site to be compared against a known database of malicious sites [27]. Detection of a malicious site can then trigger the affective feedback mechanism, delivering some form of information to the end-user. A unique log file is generated for each browser session, and records risky security behaviour triggers e.g. if a user visited a malicious site.

3 methods of affective feedback were chosen: colours, avatars and text. Previous research has indicated there are a number of types of affective feedback which could be utilised within the web browser window, to help guide users into making more appropriate security decisions. Depending on the actions of the user, they may be offered positive reinforcement because of their behaviour, negative reinforcement, or a mixture of both positive and negative.

The sentences contained in the Spengler-Zuul extension came from text in an affective word list named AFINN [28]. The avatars were chosen in relation to Ekman's 6 basic emotions [29]. Specifically, the happy and sad avatars used in this research project were drawn from work conducted by the Swiss Center for Affective Sciences [30]. Finally, colours used were chosen due to their usage in previous research projects [14][21]. The final colours chosen were: red (#CF4250), yellow (#EBA560), and green (#78BF60), producing a traffic-light system.

Multiple versions of the Spengler-Zuul extension were developed, allowing differing combinations of affective feedback to be tested against a control environment:

- Spengler-Zuul (none)- monitors users, no on-screen feedback.
- Spengler-Zuul (text)- monitors users, displays text-based affective feedback.
- Spengler-Zuul (text, avatar)- monitors users and displays text-based affective feedback, and an avatar
- Spengler-Zuul (text, colour)- monitors users and displays text-based affective feedback, and colour
- Spengler-Zuul (text, colour, avatar)- monitors users and displays text-based affective feedback, and colour. Additionally, an avatar is situated in the bottom right of the screen (Fig 1).

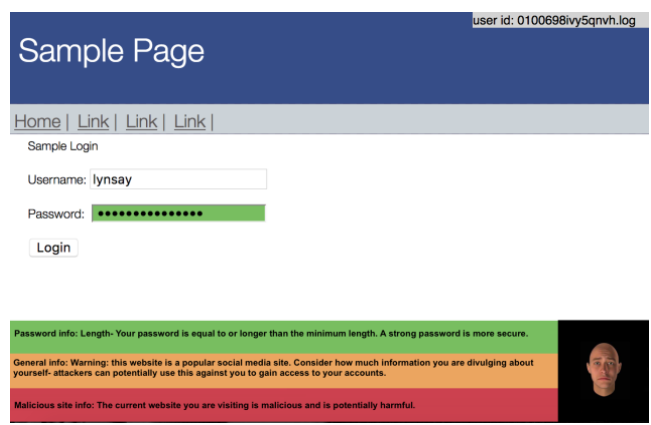

Fig 1. Screenshot of the Spengler-Zuul (text and colour and avatar) extension

*B. Experimental phase*

During the experimental process, participants were initially given an "Information For Participants" handout, noting that the experiment was testing a Firefox extension. Security awareness and behaviour were not mentioned, in relation to the experiments, in an effort to eliminate bias. Participants were then given a random USB stick, labelled with a number from 1-5 and each USB stick contained a portable version of the Firefox browser, with a version of the monitoring solution/affective feedback mechanism add-on pre-installed (Table 1). After signing the consent form, participants were asked to work their way through an instruction sheet, visiting specific websites.

Table 1. Different versions of feedback included in each experiment.

| USB Group | Feedback type | Participants (n) |
|---|---|---|
| 1 | Control | 12 |
| 2 | Text | 13 |
| 3 | Text, avatar | 16 |
| 4 | Text, colour | 14 |
| 5 | Text, colour, avatar | 17 |

Participants were asked to visit a number of predefined sites, some with false positives to trigger appropriate feedback on-screen e.g. fake malicious links to trigger warnings. During the experimental process, participants were also asked to complete a web form, asking them personal information such as hobbies. Completing this form was entirely optional however, revealing such information could have been deemed a risky security behaviour.

On completion of the computer-based part of the experiment, participants were asked to complete a paper-based questionnaire regarding how well they thought they responded to any feedback shown on-screen. In the background, the users' actions on the computer-based part of the experiment were logged, meaning the information provided in the questionnaire can be corroborated against the information in the unique log files.

IV. RESULTS

The log files gained from the monitoring solution were compared with data from the questionnaire participants answered. By comparing these approaches, an understanding of user awareness of risky security behaviour can be developed i.e. do the log files reflect what users said they actually did in the questionnaires? This multi-modal approach is comparable with work by San-José and Rodriguez [13], whereby they compared virus scan log data against questionnaire data.

The main difference Table 2 highlights is that when asked if they used a common password, participants largely said "*no*". However, there is a significant statistical difference when the log files are viewed, indicating that many users did in fact have common elements in their passwords. The same difference is seen across all experiments containing affective feedback, suggesting it did not have an impact on the actions of users in this instance.

In terms of revealing personal information, there was a significantly higher number of participants who revealed personal information about themselves (categorical information) in the log files vs. those who reported they revealed personal information in the questionnaire in experiments groups 1 (control) and 3 (text and avatar-based feedback). This potentially highlights a lack of security awareness in end users who haven't realised the level of information they divulged. This could also explain the similar results for "*Did user enter email address?*" in groups 4 (text, colour-based feedback) and 5 (text, colour, avatar-based feedback), and "*Did user visit a malicious site?*" in groups 1 (control) and 2 (text-based feedback).

Table 2. Experimental results- log files vs. questionnaire data

| Question | Group 1 (Control) | Group 2 (Text) | Group 3 (Text, avatar) | Group 4 (Text, colour) | Group 5 (Text, colour, avatar) |
|---|---|---|---|---|---|
| User revealed personal information | Yes | No | Yes | No | No |
| User entered private email address | No | No | No | Yes | Yes |
| Entered a common password | No | Yes | Yes | Yes | Yes |
| User had personal details in password | No | No | No | No | No |
| User visited a malicious site | Yes | Yes | No | No | No |

V. DISCUSSION

When the questionnaire results (report information) were compared to the log files (categorical information) there was one key question regarding risky security behaviour which produced a statistically significant result.

During the experimental process, when participants were asked if they had used a dictionary password, the majority of those asked stated "*no*". However, after analysing the requisite log files, there was a noted statistical significance which indicated that the majority of the participants had a common element in their password. The same statistical difference is noted across all of the experiments which delivered varying combinations of affective feedback.

Since similar results are seen across all experiments containing affective feedback, it suggests the delivery of the affective feedback did not have an overall impact on the actions of the participants in this instance, however, there is another potential explanation for such a result.

This result highlights there is still a need to raise security awareness in end-users and educate people regarding security behaviours which are perceived to be risky [31]. One interpretation of the result is that participants may not have been aware of the term *"dictionary word"* in relation to password. Additionally, they may not have been aware that dictionary words in passwords contribute to poor password hygiene [7].

When participants were asked if they had revealed personal information about themselves during the course of the experiments, there was a significant difference between those who reported revealing information about themselves (as per the questionnaire data), in comparison to the number of participants who categorically revealed personal information about themselves, as revealed by the appropriate log files.

In experiment 1 (control) and experiment 3 (text and avatar-based feedback) there was a significantly higher proportion of participants who categorically revealed personal information about themselves in the log files, in comparison to those who reported they revealed information about themselves when answering the questionnaire. Again, this result could be explained by the fact participants had a poor understanding of risky security behaviour, and perhaps did not understand the consequences which could arise from sharing such information.

A poor understanding of risky security behaviours could also explain the similarly statistically significant results gained when participants were asked if they entered a private email address during the course of the study. Whilst the concept of a private email address is purely subjective (what constitutes a private email address may differ depending on the user and purpose of the address), the log files were simple parsed in an effort to determine if the user had provided some form of information in the private email address field. Experiment 4 (text and colour-based feedback) and experiment 5 (text, colour and avatar-based feedback) produced statistically significant results, with more users revealing email addresses in the log files.

When asking users if they had visited a malicious website during the course of the experiment, a statistically significant result was gained in experiment 1 (control) and experiment 2 (text-based feedback). Essentially, more users categorically visited malicious sites (according to the log files) than reported visiting malicious sites in the questionnaire. Since experiment 1 does not contain any form of affective feedback whereas experiment 2 does, therefore such a result could again be attributed to the participant's lack of security awareness when browsing sites online. The proportion of those visiting malicious sites in experiment 1 also highlights the requirement for a tool to help users- if users are not provided with any feedback (like in experiment 1), they will have no way of knowing a link they are clicking on is malicious.

All information provided during the experimental process was voluntary, and this statement was clearly displayed at the top of the web pages which asked for information such as mother's maiden name, hobbies, email address, etc., which again highlights participants either chose to divulge sensitive information, or that they actively engaged in risky security behaviour by failing to read the page properly.

## VI. Conclusions

Affective feedback did not appear to have an impact on the behaviour of users as recorded by categorical information in the log files. The majority of results gained were insignificant. One anomaly was generated by experiment 5 (text, colour and avatar-based feedback) when participants were asked about the information they revealed about themselves, in comparison to the control log file. This produced a positive result, where fewer participants in experiment 5 divulged information and this suggests affective feedback may have made a difference.

However, given that all other results were insignificant, it is more plausible that the particular group of participants already possessed a good knowledge of risky security behaviours. Overall, it has been concluded affective feedback did not have an impact on participant behaviour, as per the log files.

However, the results gained still highlight an interesting point. In comparing categorical behaviour (log files) and reported behaviour (questionnaires), participants were found to have engaged in instances of risky security behaviours which they were unaware of, and this indicates a generally low level of awareness of risky security behaviour.

This research project involved a small-scale experiment. Potentially, if affective feedback was delivered over a longer period of time, on a daily basis, the log files could potentially reflect positive behavioural changes as end-users become more knowledgeable regarding the subject matter. Future work seeks to explore the long-term application of affective feedback in the domain of security awareness.